\documentclass[aps,twocolumn,,superscriptaddress,showpacs,showkeys,amsmath,amssymb,floatfix]{revtex4}
\usepackage{graphicx}
\usepackage{epsfig}
\usepackage{dcolumn}
\usepackage{bm}
\usepackage{amssymb}
\usepackage{dsfont}
\usepackage{amsmath}
\usepackage{subfigure}
\newcommand{\be}{\begin{equation}}
\newcommand{\ee}{\end{equation}}
\newcommand{\bea}{\begin{eqnarray}}
\newcommand{\eea}{\end{eqnarray}}
\newcommand{\hn}{\hat n}

\newcommand{\hD}{{\hat D}}

\newcommand{\hatm}{\hat m}
\newcommand{\tC}{{\tilde C}}

\newcommand{\tzeta}{{\tilde \zeta}}

\newcommand{\vA}{\vec A}
\newcommand{\vF}{\vec F}

\newcommand{\vX}{{\vec X}}
\newcommand{\pro}{\partial}
\newcommand{\der}{\partial}

\newcommand{\bphi}{{\mathbf \phi}}

\newcommand{\ba}{\begin{array}}
\newcommand{\ea}{\end{array}}

\newcommand{\nn}{\nonumber}

\newcommand{\uast}{\stackrel{\ast}{u}}

\begin{document}
\title{On finite energy monopole solutions in Weinberg-Salam model}
\bigskip

\author{D. G. Pak}
\affiliation{Institute of Modern Physics, Chinese Academy of Sciences,
Lanzhou 730000, China}
\affiliation{Lab. of Few Nucleon Systems,
Institute for Nuclear Physics, Ulugbek, 100214, Uzbekistan}
\author{P. M. Zhang}
\affiliation{Institute of Modern Physics, Chinese Academy of Sciences,
Lanzhou 730000, China}
\affiliation{State Key Laboratory of Theoretical
Physics, Institute of Theoretical Physics,
Chinese Academy of Sciences, Beijing 100190, China}
\author{L. P. Zou}
\affiliation{Institute of Modern Physics, Chinese Academy of Sciences,
Lanzhou 730000, China}

\begin{abstract}
We study the problem of existence of finite energy monopole
solutions in the Weinberg-Salam model starting with a most general
ansatz for static axially-symmetric electroweak magnetic fields.
The ansatz includes an explicit construction of field
configurations with various topologies described by the monopole and
Hopf charges. We introduce a unique $SU(2)$ gauge invariant
definition for the electromagnetic field. It has been proved that
the magnetic charge of any finite energy monopole solution must be
screened at far distance. This implies non-existence of finite
energy monopole solutions with a non-zero total magnetic charge. In a case
of a special axially-symmetric Dashen-Hasslacher-Neveu ansatz we
revise the structure of the sphaleron solution and show that
sphaleron represents a non-trivial system of monopole and antimonopole
with their centers located in one point.
This is different from the known interpretation of the
sphaleron as a monopole-antimonopole pair like Nambu's
"dumb-bell". In general, the axially-symmetric magnetic field may
admit a helical structure. We conjecture that such a solution
exists and estimate an upper bound for its energy, $E_{bound}=4.65$ TeV.
\end{abstract}
\maketitle

\section{Introduction}

One of puzzling questions in modern high energy physics
is whether monopoles exist in realistic theories of fundamental
interactions. The well known singular monopoles
\cite{dirac,wuyang,nambu77} can not represent physical observable objects.
Discovery of such solely existing monopoles of pure electromagnetic (or gravitational)
nature would break the present fundamental laws of physics.
In this regard, composite monopole solutions described within
the framework of the standard electroweak model could
be more favorable candidates for monopoles expected to be found at LHC \cite{CERN,IJMPA}.
An important issue in search of monopoles is to provide strong theoretical
reasons for their existence. So far known solutions either lack a regular structure
or need essential extension beyond the standard model \cite{thooft,polyakov74,chomaison}.

Yang-Mills-Higgs (YMH) theory with a complex doublet Higgs field
does not admit finite energy monopole solutions. The reason is
that finite energy Higgs field configurations in the asymptotic
region at space infinity forms a three-dimensional sphere. Since a
two-dimensional sphere is contractible on $S^3$ the relative
homotopy group $\pi_2(S^3,S^2)=\pi_2(S^3)$ is trivial and can not
provide the topological monopole charge. It has been observed
\cite{chomaison} that in Weinberg-Salam model the situation
changes drastically due to presence of the hypermagnetic gauge
symmetry $U_Y(1)$ which
implies existence of monopole solutions. Indeed,
an example of such a solution is given by a singular Cho-Maison
monopole \cite{chomaison}. The question whether finite
energy monopole solutions exist in the Weinberg-Salam model has not been
resolved so far and represents an important issue.

In the present paper we undertake a systematic study of possible
monopole solutions in the Weinberg-Salam model. Our consideration is
restricted by the case of static axially-symmetric magnetic field
configurations. A general ansatz for axially-symmetric solutions
in the electroweak theory was suggested in \cite{brihkunz94}. We
propose an alternative general axially-symmetric ansatz based on
the formalism of a gauge invariant decomposition
\cite{choprd80,choprd81,duan} in a natural basis frame $\hat m_i$
($i=1,2,3$) in the internal space of the group $SU(2)$. The natural
basis frame is determined by $SU(2)$ triplet vector field $\hat m
\equiv \hat m_3$ constructed from the Higgs field. A key point in
our ansatz is an explicit parametrization of the vector field
$\hat m$ in terms of two arbitrary functions which determine the
topology of the Higgs field described by two topological
invariants. Our approach allows to describe topology of the Higgs
and gauge bosons and trace the topological origin of magnetic like
solutions. Applying our ansatz and finite energy condition one can
find all possible local solutions for the gauge fields and Higgs
boson in the asymptotic region at space infinity.
{\it For a wide class of static
axially-symmetric field configurations we have proved that any
possible finite energy monopole solution must have a totally
screened magnetic charge}. We show that finite energy monopole
solutions with a non-zero total magnetic charge do not exist in
the Weinberg-Salam model. Possible finite energy magnetic solutions
can be represented by monopole-antimonopole systems or pure
magnetic field configurations only with vanishing total magnetic
charge. We will consider simple examples of such two types of
magnetic solutions.

In a special case of axially symmetric magnetic fields with
vanishing azimuthal magnetic field component
the general axially symmetric ansatz reduces to
Dashen-Hasslacher-Neveu (DHN) ansatz \cite{DHN, manton83}. Within
the DHN ansatz various sphaleron solutions were obtained in the
Yang-Mills-Higgs theory and Weinberg-Salam model
\cite{DHN,manton83,klinkmanton84,klink90,brih92,brih93a,brih93b}.
It was suggested to interpret the original DHN sphaleron as a
monopole-antimonopole pair \cite{hind94,hind94b,volkov}. The
interpretation was conditioned by use of a special definition for
the electromagnetic field tensor which supposed to have
ambiguity in its definition \cite{hind94,hind94b,colemanasp}. One
should notice, a physical concept and the respective mathematical
definition for the electromagnetic field as a physical observable
quantity should be unique and invariant under $SU(2)$ gauge
transformation. We show that the electromagnetic vector potential can
be uniquely defined in $SU(2)$ gauge invariant manner. With this
we revise the internal structure of the sphaleron and demonstrate
that DHN sphaleron represents a monopole-antimonopole pair with
screened monopole and antimonopole charges. Our view is different
from Nambu's "dumb-bell" monopole-antimonopole interpretation
\cite{hind94,hind94b,volkov}.

In general, the magnetic field may admit non-vanishing helicity.
We consider a simple possible magnetic solution in
Yang-Mills-Higgs theory which possesses only the azimuthal
non-vanishing magnetic flux. Applying variational method we obtain
an estimate for the upper energy bound, $E_{bound}=4.65$ TeV.

The paper is organized as follows. In Section II a general ansatz
for static axially-symmetric magnetic fields is proposed. In
Section III the problem of existence of finite energy monopole
solutions is studied. Internal structure of the sphaleron solution
is revised in Section IV. We consider a new type of magnetic like
solution with a non-vanishing azimuthal magnetic flux in Section V.
First we find an exact numeric magnetic solution in a simple
$CP^1$ model reduced from the Yang-Mills-Higgs theory. Then we
study the structure of a similar solution in a full Yang-Mills-Higgs
theory by using variational methods.

\section{A general ansatz for static axially symmetric magnetic field}

The bosonic part of the Weinberg-Salam model is described by the
following Lagrangian ($\mu,\nu=0,1,2,3$, $a,b=1,2,3$)
\bea
&& {\cal L} =
-\dfrac{1}{4} ({\vec F}_{\mu\nu})^2-\dfrac{1}{4} (G_{\mu\nu})^2
  - |D_\mu \bphi|^2-\dfrac{\lambda}{2} ({\bphi}^\dagger {\bphi} -
 \dfrac{v^2}{2})^2, \nn \\
&& {F}_{\mu\nu}^a=\pro_\mu A_\nu^a-\pro_\nu A_\mu^a +g \epsilon^{abc} A_\mu^b A_\nu^c, \nn \\
&&G_{\mu\nu}=\pro_\mu B_\nu - \pro_\nu B_\mu, \nn \\
&&D_\mu=\pro_\mu -\dfrac{ig}{2} \vec \sigma \cdot \vec A_\mu -i \dfrac{g'}{2} B_\mu, \label{LagrWS}
\eea
where $\vA_\mu$ and $B_\mu$ are the gauge fields
corresponding to the electroweak gauge group $SU(2)\times U_Y(1)$,
and $\bphi$ is the Higgs complex scalar doublet.

The equations of motion have the following  form
\bea
D^\nu \vec F_{\nu\mu}&=&\dfrac{ig}{2}\Big ( \bphi^\dagger \vec \sigma (D_\mu \bphi)
-(D_\mu \bphi)^\dagger \vec \sigma \bphi \Big ), \nn \\
\der^\nu G_{\nu\mu}&=&\dfrac{ig'}{2}\Big ( \bphi^\dagger (D_\mu \bphi)
-(D_\mu \bphi)^\dagger \bphi \Big ), \nn \\
D^\mu D_\mu \bphi&=& \lambda (\bphi^\dagger \bphi-\dfrac{v^2}{2})^2. \label{wseqs}
\eea

To construct a most general ansatz for static axially-symmetric magnetic
field configurations we apply a gauge invariant
decomposition of the gauge potential
in arbitrary orthonormal  frame $(\hat n_1,\hn_2,\hn_3)$
in the internal space of $SU(2)$ \cite{choprd80,choprd81}
\bea
\vA_\mu&=&\hat A_\mu +\vX_\mu,  \nn \\
\hat A_\mu&=&A_\mu \hn+\vec C_\mu,  \nn \\
\vec C_\mu&=&-\dfrac{1}{g} \hn \times \pro_\mu \hn,~~~~~~~
\vX_\mu= X_\mu^{1,2} \hat n_{1,2}, \label{abeldec}
\eea
where $\hat A_\mu $ is a restricted gauge potential,
$\vX_\mu$ contains two off-diagonal components of the gauge potential,
$\hn\equiv \hn_3$ is a basic $SU(2)$ vector field
which determines the basis frame $\hn_i$ $(i=1,2,3)$ in the group $SU(2)$.
The vector $\hn$ satisfies the covariant constance condition
\bea
\hat D_\mu \hat n \equiv (\pro_\mu \hn+g \hat A_\mu \times \hn)=0.
\eea
For any given unit vector field $\hat n$ the full $SU(2)$ gauge
field strength can be decomposed into the Abelian and off-diagonal parts
in a gauge invariant manner
\bea
\vF_{\mu\nu}&=&(F_{\mu\nu}+H_{\mu\nu}) \hn \nn +\\
 &&\hD_\mu \vX_\nu-\hD_\nu \vX_\mu+g \vX_\mu \times \vX_\nu, \nn \\
 F_{\mu\nu}&=& \pro_\mu A_\nu-\pro_\nu A_\mu, \nn \\
H_{\mu\nu}&=&\dfrac{1}{g}\epsilon^{abc}\hn^a \der_\mu \hn^b \der_\nu \hn^c=
               \pro_\mu \tC_\nu-\pro_\nu \tC_\mu, \label{cdual}
\eea
where the Abelian dual magnetic potential $\tC_\mu$
is defined by the vector field $\hn$ (up to dual Maxwell type $\tilde U(1)$ gauge transformation)
\cite{choprd81}.

The vector field $\hat n$
can be expressed in terms of the complex function $u(x)$
or in terms of the complex projective coordinates $\zeta_{1,2}$
on the sphere $S^2$ by using a standard
stereographic projection
\bea
\hat n &=& \dfrac{1}{1+u \uast}
 \left (\ba{c}
  u+\uast\\
  -i (u-\uast)\\
 u \uast-1\\
            \ea
           \right ), \nn \\
u &=&\dfrac{\zeta_1}{\zeta_2},~~~~~~~~ \hn= \zeta^+ \vec \sigma \zeta, \label{nstereo}
\eea
where $\zeta_{1,2}$ form a unit complex $SU(2)$ vector doublet.

We introduce an ansatz for
the $SU(2)$ vector field $\hat n$ which can be written
for the complex function $u(x)$ in spherical coordinates as follows
\bea
u(r,\theta,\varphi)&=& e^{-i m \varphi} \big (\cot(\dfrac{n\theta}{2}) f(r,\theta)+
i \csc (\dfrac{n\theta}{2})Q(r,\theta)\big), \nn \\
&&\label{ansfQ}
\eea
where $(m,~n)$ are integer winding numbers which specify
the topological monopole charge, $Q_m$, and the
Hopf charge, $Q_H$, of the magnetic field configuration $H_{mn}$
\bea
Q_m&=&\dfrac{1}{A(S)}\int_{S^2} H_{ij} \cdot d \sigma^{ij}, \nn \\
Q_H&=& \dfrac{1}{32 \pi^2}\int d^3x \epsilon^{ijk} \tC_i H_{jk} , \label{monHopfch}
\eea
where $A(S)$ is the surface area of a sphere $S^2$.

Let us consider a parametrization for the Higgs field
suitable for description of its topological properties.
One can parameterize the Higgs field in terms of a
scalar field $\rho(x)$ and a unit complex $SU(2)$ doublet $\tzeta$
with explicit extracting the $U_Y(1)$ exponential factor containing a
field variable $\omega(x)$
\bea
&&\bphi= \dfrac{1}{\sqrt 2} \rho \tzeta e^{i\omega(x)},  \nn \\
&& \tzeta^+ \tzeta=1. \label{higgspar}
\eea
In the following we
will use a gauge condition $\omega(x)=0$.
One can define a real
$SU(2)$ triplet vector field $\hat m$ constructed directly from
the Higgs field
\bea
\hat m^a&=&\hat \phi^+ {\vec \sigma}^a \hat \phi=\tzeta^+ {\vec \sigma}^a \tzeta, \nn \\
\hat \phi&=& \dfrac{\phi}{|\phi|},~~~~~ \hat m^2=1, \label{hatm}
\eea
where $\vec \sigma^a$ are Pauli matrices.

The vector field $\hn$ in Abelian decomposition (\ref{abeldec})
does not possess its own equations of motion and contains only topological degrees
of freedom. Another feature of the gauge invariant Abelian decomposition
is appearance of two types of $SU(2)$ gauge symmetries, so-called
"active" and "passive" ones \cite{chopak02}. Respectively,
the vector $\hn$ transforms in different ways under these two types
of gauge transformations.
Contrary to this, the vector $\hat m$,
(\ref{hatm}), is determined completely by a given Higgs field.
So the gauge transformation law for $\hat m$ is fixed uniquely,
and the basis frame $\hat m_i$ constructed from $\hat m_3\equiv \hat m$ represents
a natural basis in the decomposition of the gauge potential.
In general one can perform decomposition of the gauge potential
in various basis frames defined by any vector field $\hn$.
To fix the arbitrariness of choosing $\hn$
one should impose a constraint on $\hn$ and $\hat m$.
In practical applications it is convenient to
fix the vector $\hn$ using one-to-one or
double covering mapping between two spheres defined by the
vectors $\hn$ and $\hat m$.

For the gauge potentials $A_\mu,~X_\mu^{1,2}$
we adopt a most general form given by arbitrary functions
depending on two spherical coordinates $(r,\theta)$.
In the case of static magnetic fields one can choose a temporal gauge $\vA_0=0$.
It is convenient to parameterize the Higgs field
in terms of three independent functions $\beta_{k}(r,\theta)$ ($k=1,2,3$)
\bea
\bphi&=&\dfrac{1}{\sqrt 2}(\beta_1+i \beta_3, e^{im\varphi} \beta_2).
\eea
In some cases it is suitable to change further the variables and
express $\beta_{k}$ in terms of a Higgs real scalar
field $\rho$ and two angle functions $\tilde T(r,\theta),\tilde S(r,t)$
\bea
\beta_1(r,\theta)&=& \rho(r,\theta) \cos \dfrac{\tilde T(r,\theta)}{2} \sin \dfrac{\tilde S(r,\theta)}{2}, \nn \\
\beta_2(r, \theta)&=& \rho(r,\theta) \sin \dfrac{\tilde T(r,\theta)}{2} \sin \dfrac{\tilde S(r,\theta)}{2}, \nn \\
\beta_3(r,\theta)&=& \rho(r,\theta) \cos \dfrac{\tilde S(r,\theta)}{2}. \label{beta123}
\eea
Note that parametrization (\ref{beta123}) with the functions $\beta_{1,2,3}$
corresponds in general to topology of a two-dimensional sphere $S^2$ for
a constant valued Higgs scalar field
\bea
\beta_1^2+\beta_2^2+\beta_3^2=\rho^2. \label{beta3}
\eea
Our ansatz includes totally fifteen field variables which
satisfy fifteen equations of motion.
The functions $\beta_{1,2,3}$ are treated as initial independent
variables of the Higgs field,
and in the following we will always use the original Weinberg-Salam equations
obtained by variation of the Lagrangian (\ref{LagrWS})
with respect to the fields $B_\mu, \vec A_\mu, \beta_{1,2,3}$.

To verify that our ansatz leads to axially symmetric
configurations let us consider the structure of the Lagrangian (\ref{LagrWS}).
It has been shown in \cite{choprd81} that within the formalism of
gauge invariant Abelian projection the Yang-Mills part in the Lagrangian (\ref{LagrWS})
describes a theory of a charged matter field $X_\mu=\frac{1}{\sqrt 2}(X_\mu^1+i X^2_\mu)$
interacting with an Abelian gauge field $A_\mu+\tC_\mu$.
The original $SU(2)$ gauge transformation for the gauge fields
can be written in the form \cite{choprd81,chopak02}
\bea
\delta \hat n&=&-\vec \alpha \times \hat n, \nn \\
\delta A_\mu&=& \dfrac{1}{g} \hn \cdot \pro_\mu \vec \alpha, \nn \\
\delta \vX_\mu &=& -\vec \alpha \times \vX_\mu,   \label{transhn}
\eea
which implies that the vector field $\vX_\mu$ transforms covariantly,
i.e., it behaves as a matter field.
By direct calculation one can check that ansatz (\ref{ansfQ}) describes
an axially symmetric
configuration for the magnetic field $H_{\mu\nu}$.
So that the axially symmetric form of the gauge potentials $A_\mu, X_\mu^{1,2}$
guarantees axially-symmetric configurations for all fields
and consistence with the equations of motion as well.

Let us now consider definitions of gauge invariant quantities in
the Weinberg-Salam model.
The hypermagnetic field strength tensor $G_{\mu\nu}$
is gauge invariant due to the Abelian structure of the gauge group $U_Y(1)$.
For the $SU(2)$ gauge potential $\vec A_\mu$ we apply
a gauge invariant decomposition in
the natural basis frame $(\hat m_1, \hat m_2, \hat m_3\equiv \hat m)$.
Since the vector field $\hatm$ belongs to an adjoint representation
of the group $SU(2)$,
the Abelian projection onto $\hatm$ direction provides
a field tensor which is invariant under $SU(2)$ gauge transformation
\bea
F^{full}_{\mu\nu}&=&\vF_{\mu\nu} \hatm \nn \\
  &=&F_{\mu\nu}+H_{\mu\nu}+g X_\mu^1 X_\nu^2-g X_\nu^1 X_\mu^2. \label{emthn}
\eea
An important issue of the gauge invariant decomposition
is that for a given covariant vector $\hat m$ one can
define an Abelian gauge vector potential ${\cal A}_\mu$
of the Maxwell type which is gauge invariant
under arbitrary $SU(2)$ transformation \cite{choprd80,choprd81}
\bea
{\cal A}_\mu=A_\mu+\tilde C_\mu. \label{calA}
\eea
This allows to define a corresponding  $SU(2)$ gauge invariant
Abelian field strength
\bea
{\cal F}_{\mu\nu}&=&F_{\mu\nu}+H_{\mu\nu}. \label{emtthooft}
\eea
The definition coincides with the 't Hooft-Polyakov magnetic field tensor
in YMH theory \cite{thooft,polyakov74}.
With this we can introduce a unique gauge invariant
definition for the electromagnetic gauge potential and neutral gauge boson
\bea
&&A^{em}_\mu=\cos \theta_W  B_\mu + \sin \theta_W {\cal A}_\mu, \nn \\
&&Z_\mu=-\sin \theta_W B_\mu + \cos \theta_W {\cal A}_\mu. \label{Aem}
\eea
One should stress, that for any given Higgs field configuration
the expressions for $A^{em}_\mu, Z_\mu$ are invariant under
$SU(2)$ transformation and do not depend on a specific choice of
a gauge condition for $\hat m$. In a unitary gauge, $\hat m=(0,0,1)$,
or equivalently $\tzeta=(0,1)$, the
expressions for $A^{em}_\mu,Z_\mu$ reduce to the standard definitions
of the electromagnetic potential and neutral gauge boson in the Weinberg-Salam model.
A respective $SU(2)$ gauge invariant electromagnetic field tensor
is given by
\bea
&&F^{em}_{\mu\nu}=\cos \theta_W G_{\mu\nu} + \sin \theta_W {\cal F}_{\mu\nu}. \label{Fem}
\eea

An alternative definition for the electromagnetic field
with  using a full Abelian field tensor $F^{full}_{\mu\nu}$ instead of ${\cal F}_{\mu\nu}$
was suggested in \cite{hind94,hind94b}.
However, a corresponding definition for the electromagnetic vector potential
contains an $SU(2)$ gauge non-invariant term $A_\mu^a \hat \bphi^a$,
and, in addition, the Bianchi identities for the
electromagnetic field are not fulfilled anymore.
So, such a definition can not be accepted in a consistent manner.

The presence of the Abelian gauge potential ${\cal A}_\mu$
allows to define an $SU(2)$ gauge invariant monopole charge,
${\cal Q}_m$, and an
analog of the Chern-Simons number, ${\cal Q}_{CS}$,
\bea
{\cal Q}_m&=&\dfrac{1}{A({\cal S})}\int_{{\cal S}^2} {\cal F}_{ij} \cdot d \sigma^{ij}, \nn \\
{\cal Q}_{CS}&=& \dfrac{1}{32 \pi^2}\int d^3x \epsilon^{ijk} {\cal A}_i {\cal F}_{jk}, \label{gimonHopfch}
\eea
where $A({\cal S})$ is the area of a closed two-dimensional surface ${\cal S}^2$.

Let us consider a reduction of the general axially-symmetric ansatz
to Dashen-Hasslacher-Neveu ansatz.
In Abelian decomposition
of the gauge potential $\vec A_\mu$ we use the following parametrization
for the vector $\hn$ in terms of one function $f(r,\theta)$,
or angle function $T(r,\theta)$ equivalently
\bea
\zeta_1&=&\cos \frac{T(r,\theta)}{2}=\dfrac{\cos \frac{n\theta}{2} f(r,\theta)}
                     {\sqrt {\cos^2 \frac{n\theta}{2} f^2(r,\theta)+\sin^2
            \frac{n\theta}{2}}}, \nn \\
\zeta_2&=& e^{im\varphi} \sin\frac{T(r,\theta)}{2}.  \label{zetaT}
\eea
The parametrization (\ref{zetaT}) corresponds to a special ansatz
(\ref{ansfQ}) when $Q(r,\theta) \equiv 0$.
For the Higgs field we apply a reduced parametrization
(\ref{beta123}) with the constraints
\bea
\tilde S(r,\theta)&=&\pi, ~~~~~\tilde T(r,\theta)=p T(r,\theta), \label{pT}
\eea
where the last relationship establishes a connection between the
vectors $\hn $ ($\zeta$) and $\hat m$ ($\tzeta$), $p$ is an integer number corresponding
to cover mapping of the sphere $S^2$. Finally,
a simple setting
\bea
&& A_r^3=A_\theta^3=A_r^1=A_\theta^1=A_\varphi^2=0, \nn\\
&& A_r^2=K_1,~~~A_\theta^2=K_2, \nn \\
&& A_\varphi^3=K_3,~~~A_\varphi^1=K_4,    \label{DHNK}
\eea
leads to a modified DHN axially symmetric ansatz \cite{DHN,manton83}
with four functions $K_{1,...,4}$ and two functions $\beta_{1,2}$
(or $(\rho, \tilde T)$). Notice, equations of motion are obtained by variation
of the initial Lagrangian with respect to independent variables
$K_{1,...,4}, \beta_{1,2}$.
In comparison with the original DHN ansatz one has an additional field degree of freedom
represented by the function $f(r,\theta)$.
A simplest choice $f(r,\theta)=1$ leads to the original DHN ansatz.
Note thata in the case of a static magnetic field the DHN ansatz implies that
the azimuthal
magnetic field component ${\cal F}_{r\theta}$ vanishes identically.
So, within the DHN ansatz one can not describe magnetic field configurations
with a helical structure. In addition, the relationship (\ref{beta3}) reduces to
the following one
\bea
\beta_1^2+\beta_2^2=\rho^2
\eea
which constrains the topology of the Higgs field configuration
to a foliation $R^+\times S^1$, $R^+$ is a half-line corresponding to positive values of
the Higgs scalar $\rho$.

For further numeric purpose it is suitable
to write down the energy functional corresponding to the Weinberg-Salam
Lagrangian (\ref{LagrWS}) for static magnetic field configurations in terms of
dimensionless variables
$ r \rightarrow r m_W, \,\,
 k \equiv  2 \sqrt \lambda /g = m_H/m_W ,\,\,
\vA_\mu\rightarrow \vA_\mu g/m_W,\,\,\tC_\mu\rightarrow \tC_\mu g/m_W,
\,\,B_\mu\rightarrow B_\mu g'/m_W,\,\, \rho \rightarrow \rho/v $.
In these variables the energy functional takes the following form ($m,n,=1,2,3$)
\bea
E &=& \dfrac{m_W}{g^2}  \int d^3 x \Big [\dfrac{1}{4} {\vec F_{mn}}^2
                     +\dfrac{1}{4}\kappa {\vec G_{mn}}^2 + \nn \\
           &&  2 |D_m \bphi|^2 + \dfrac{k^2}{2} (\rho^2-1)^2 \Big ], \label{energy}
\eea
where $\kappa = \dfrac{g^2}{g'^2}=3.324$,
$k^2=\dfrac{m_H^2}{m_W^2}=2.441$ and
$m_W =80.385 $ Gev , $m_H=125.6 $ Gev, $\, \sin^2\theta_W=0.23126$. Numeric value of the
mass factor in front of the integral is $\dfrac{m_W}{g^2}=203$ Gev.

\section{Non-existence of finite energy monopoles with unscreened magnetic charge}


The form of the Higgs potential in the Weinberg-Salam model
implies that finite energy Higgs field configurations
at space infinity must satisfy the condition
$|\bphi|^2=1$ which describes a three-dimensional sphere $S^3$.
This implies that topology
of the Higgs field in the asymptotic region is determined by one non-trivial
homotopy group $\pi_3(S^3)=Z$.
Topological classes of $SU(2)$ gauge field are described by the same homotopy group
since the gauge group manifold $SU(2)$ is isomorphic to a sphere $S^3$.
Absence of the non-trivial second homotopy $\pi_2(S^2)$
serves as a strict argument against existence of finite
energy monopole solutions in YMH theory with a
complex Higgs field.
In the Weinberg-Salam model the hypermagnetic symmetry
can be fixed in a consistent manner with the standard unitary gauge
by imposing a constraint $\omega=0$ in (\ref{higgspar}).
This leads to appearance of additional two non-trivial
homotopy groups $\pi_{2,3}(S^2)=Z$
which provide necessary conditions for
monopole and knot solutions in the theory \cite{chomaison,choknot,pakknot}.
However, the problem of existence of such solutions with a finite energy
has not been studied carefully so far.
We consider this problem by studying
local solutions near the space infinity and taking into account
the finite energy condition.

In the asymptotic region near space infinity
the structure of monopole solutions is determined by
behavior of the radial magnetic field component ${\cal F_{\theta\varphi}}$,
other magnetic field components turn into pure gauge configurations.
So the DHN ansatz can be applied properly in the analysis
of solutions in the asymptotic limit. For the case of
the Weinberg-Salam model the DHN ansatz can be consistently extended by
adding a hypermagnetic field component $B_\varphi$,
all other components must be identically vanished.

Imposing conditions $T(r,\theta)=n \theta$, i.e., $f(r,\theta)=1$,
and setting the parameter values $(m=1,~p=1,2,~n=1,2)$
we can apply the DHN ansatz (\ref{zetaT}-\ref{DHNK})
and find regular monopole like solutions.
For other values of $(p,~n)$ we didn't find regular solutions.
With this an explicit expression for
the gauge invariant magnetic field ${\cal F}_{mn}$
can be written as follows
\bea
{\cal F}_{r\theta}&=&0, \nn \\
{\cal F}_{r\varphi}&=& K_1 \sin(n\theta)+\pro_r K_3-K_4 K_1, \nn \\
{\cal F}_{\theta\varphi}&=&K_2 \sin(n\theta)+\pro_\theta K_3-K_4 (K_2+n). \label{Ftf}
\eea
In addition we impose a gauge condition $K_1(r,\theta)=0$ due to presence of
the residual gauge symmetry \cite{DHN,klinkmanton84,KKprd}.

Let us consider possible solutions with a set of parameters
$(n=1,~p=1)$. For simplicity we restrict our consideration by
the limiting case $\lambda=0$. To find proper boundary conditions for the
gauge field components
$B_\varphi,K_{2,3,4}$ one can simplify the asymptotic expression
for the energy functional (\ref{energy}) in the lowest order
of series expansion to the following form
\bea
&&E=\int dr d\theta d\varphi {\cal E}^{asym} ,\nn \\
&&{\cal E}^{asym}=\Big (B_\varphi^{inf}(\theta)+(K_3^{inf}(\theta)
+\cos\theta-1) \Big)^2+ \nn \\
&&\Big (K_4^{inf}(\theta)-\sin\theta \Big )^2+\Big (K_2^{inf}(\theta)+1 \Big )^2 \sin^2\theta + \nn \\
&&~~O\Big (\dfrac{1}{r^2}\Big ),
\eea
where the asymptotic energy density ${\cal E}^{asym}$
includes the integration volume $r^2 \sin \theta$, and
$B_\varphi^{inf}(\theta),~K_{2,3,4}^{inf}(\theta)$ are asymptotic
functions for the respective gauge fields at space infinity.
Finite energy condition implies the following constraints
\bea
&&K_3^{inf}(\theta)=1-\cos\theta -B_\varphi^{inf}(\theta), \nn \\
&&K_2^{inf}(\theta)= -1, \nn \\
&&K_4^{inf}(\theta)=\sin\theta. \label{finconds}
\eea
In a weak coupling limit, $g'=0,~B_m=0$, the finite energy condition
leads to vanishing the radial
magnetic field component ${\cal F}_{\theta\varphi}$ (\ref{Ftf})
in asymptotic region in agreement with the fact of absence of monopole
solutions in pure YMH theory.
For the case $g' \neq 0$ the structure of asymptotic solutions is more
rich. Substituting the asymptotic functions (\ref{finconds}) into all equations of motion
one can verify that in lowest order approximation
all equations are satisfied except equations
for $B_\varphi$ and $K_3$. The full equation of motion for the hypermagnetic field
reads
\bea
&&-2 \Big[ B_\varphi-1+\cos \theta +K_3 \Big] \rho^2 +\nn \\
&&\kappa \Big [r^2 \pro_{rr} B_\varphi+ \pro_{\theta\theta} B_\varphi-
\cot \theta \pro_\theta B_\varphi \Big] =0. \label{fullBf}
\eea
The equation in asymptotic region reduces to two equations enclosed in square brackets.
The first one gives the same relationship between $B_\varphi^{inf}$ and $K_3^{inf}$ as in (\ref{finconds}),
and the second one has a simple solution
\bea
B_\varphi^{inf}(\theta)=C_1 +C_2 \cos \theta,  \label{diracmon}
\eea
where $C_1, C_2$ are integration constants.
The equation for $K_3^{asym}$ is the same as for $B_\varphi^{inf}$, and
it gives the following solution consistent with finite energy conditions (\ref{finconds})
\bea
K_3^{inf}(\theta)=1-C_1-(1+C_2) \cos \theta.
\eea
With this a non-zero component of the Abelian gauge invariant
magnetic field tensor takes the form
\bea
{\cal F}_{\theta\varphi}=C_2 \sin\theta
\eea
which implies a finite magnetic flux through the sphere of infinite radius,
i.e., a non-zero magnetic charge.
The energy density in the asymptotic region has
Wu-Yang monopole like behavior
\bea
{\cal E}^{asym}=\dfrac{C_2^2}{2r^2} (1+\kappa) \sin \theta.
\eea

In a similar manner one can find asymptotic solutions
for the gauge fields in the case $p=1,2$, $n=1,2$
\bea
&&B_\varphi^{inf}(\theta)=C_1 +C_2 \cos \theta, \nn \\
&&K_2^{inf}(\theta)= -pn, \nn \\
&&K_3^{inf}(\theta)=-\cos(n \theta)-(B_\varphi^{inf}(\theta)-1)\cos(n(p-1)\theta), \nn \\
&&K_4^{inf}(\theta)=\sin(n\theta)-(B_\varphi^{inf}(\theta)-1)\sin(n(p-1)\theta). \nn \\
&& \label{fincondpn}
\eea
A corresponding asymptotic expression for the non-vanishing component of the
gauge invariant Abelian magnetic field is given by
\bea
{\cal F}_{\theta\varphi}&=&-\pro_\theta B_\varphi^{inf}(\theta).
\eea
The asymptotic energy density has the same form for various sets of parameter
values $(p=1,2,~n=1,2)$
\bea
{\cal E}^{asym}=\dfrac{C_2^2}{2r^2} (1+\kappa) \sin \theta.
\eea
As expected, the expressions for gauge invariant quantities like the energy density and
magnetic field ${\cal F}_{mn}$ do not depend on a choice of the parameter values $(p,n)$.
Now it becomes clear that the non-vanishing contribution
to the magnetic flux created by $F^{em}_{\theta\varphi}$
originates only from the hypermagnetic gauge potential
$B_\varphi$. For non-zero values of the integration constant
$C_2$ the hypermagnetic field
creates a non-vanishing magnetic flux through the sphere of infinite radius.
This implies that one has a divergence of the hypermagnetic
field at least in some point inside the sphere.
Since the hypermagnetic magnetic field is gauge invariant and
divergenceless,
the magnetic flux through any closed two-dimensional surface must vanish,
otherwise the magnetic field will possess a singularity which will lead
to infinite total energy. Note that the energy functional
(\ref{energy}) contains four terms each of them is positively defined.
Due to this an infinite energy contribution of the hypermagnetic field
can not be compensated by contributions of the Higgs boson or
 $SU(2)$ gauge fields.
We conclude, {\it for any static axially-symmetric finite energy monopole-like solutions
the magnetic charge must be screened at far distance. The hypermagnetic field can not
serve as a source of the monopole with a non-zero magnetic charge
localized inside any two-dimensional
closed surface.} In other words, for any possible finite energy monopole solutions the
magnetic flux through the sphere of infinite radius must vanish,
and the magnetic flux of the hypermagnetic field through any closed surface
of a finite radius must be zero as well. This implies immediately
non-existence of a finite energy solution representing
a system of monopoles and antimonopoles localized in different points.
In particular, a finite energy Nambu type of monopole-antimonopole
does not exist at least for the case of static non-rotating monopole-antimonopole pair.

\section{Sphaleron as a monopole-antimonopole pair with screened magnetic charges}

Let us consider the case $f=1, Q=0$, (\ref{ansfQ}),
when the general axially-symmetric ansatz
reduces to the original DHN ansatz.
In the limit of small coupling constant $g'$ the Weinberg-Salam
Lagrangian for the bosonic fields turns into the Lagrangian
of Yang-Mills-Higgs theory by simple setting $B_\mu=0$.
So that, solutions of the YMH  theory represent
approximate solutions of the Weinberg-Salam model in lowest order
of perturbation theory. In this section we constrain our study by
the case of YMH theory in the limit $\lambda=0$.
Our goal is to study all possible monopole like solutions
within the DHN ansatz, (\ref{zetaT}-\ref{DHNK}).
With setting $m=1,~T(r,\theta)=n\theta, \tilde T(r,\theta)=pn \theta$
one can find the following
reduction ansatz in the case of parameter values $(p=2,n=1)$
\bea
K_1&=&0, \nn \\
K_2&=&-1+K(r), \nn \\
K_3&=&0, \nn \\
K_4&=&(1-K(r))\sin \theta, \nn \\
\tilde T&=& 2 \theta. \label{sphI}
\eea
One can easily verify that ansatz (\ref{sphI}) with a radial trial function $K(r)$
leads to the known DHN sphaleron solution \cite{DHN,klinkmanton84}.
In a similar manner one can find a solution
applying the DHN ansatz with parameter setting $(p=2,n=2)$
\bea
K_1&=&0, \nn \\
K_2&=&-3+K(r), \nn \\
K_3&=&(1+K(r)) \sin^2\theta, \nn \\
K_4&=&\dfrac{1}{2}(3-K(r))\sin(2 \theta), \nn \\
\tilde T&=& 4 \theta. \label{sphII}
\eea
The Higgs field $\rho(r)$ and the function $K(r)$ have only radial dependence,
but the whole Higgs complex scalar doublet and gauge fields have
axially-symmetric configuration.
Another one axially-symmetric solution can be obtained by
using the ansatz with the set of parameters $(p=1,n=2)$
\bea
K_1&=&0, \nn \\
K_2&=&-3-K(r), \nn \\
K_3&=&(1-K(r)) \sin^2\theta, \nn \\
K_4&=&\dfrac{1}{2}(3+K(r))\sin(2 \theta), \nn \\
\tilde T&=& 2 \theta. \label{sphIII}
\eea
Parametrization of the functions $K_{2,3,4}$ in terms of the trial
radial function $K(r)$
is chosen in such a way that
final differential equations and boundary conditions for the functions
$K,\rho$ are the same for all three solutions.
Direct substituting the ansatz into the
equations of motion reduces all equations to two
ordinary differential equations
\bea
&&r^2 K''=K(K^2-1)+r^2 \rho^2 (K+1), \nn \\
&&r^2 \rho''+2 r \rho'=\dfrac{1}{2}\rho(1+K)^2.
\eea
The energy density reduces to a simple expression
\bea
{\cal E}=\dfrac{1}{r^2} \Big(2 r^2 \rho'^2+ K'^2+\dfrac{(1-K^2)^2}{2 r^2}+
\rho^2(1+K)^2\Big).
\eea
So that one has three gauge equivalent representations for the DHN sphaleron.
To solve the equations one choose boundary conditions
consistent with the finite energy constraint
\bea
K(r=0)=1, ~~~~~~~&&K(r=\infty)=-1, \nn\\
\rho(r=0)=0,~~~~~~~&& \rho(r=\infty)=1.
\eea
Notice, the boundary values for the function $K(r)$ correspond
to vacuum configurations for $SU(2)$ gauge potential near the
origin and in the asymptotic region at space infinity.
Numeric solution is depicted in Fig. 1.
\begin{figure}[htp]
\centering
\includegraphics[width=60mm,height=50mm]{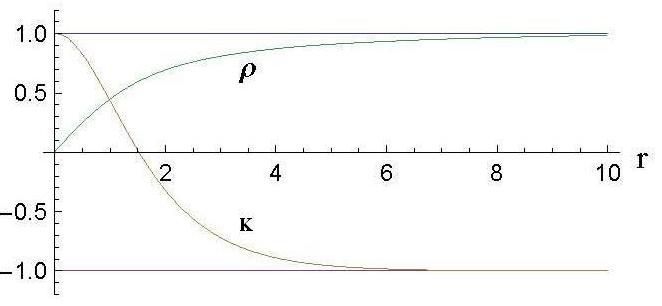}
\caption[plot1]{Solution for the functions $K,\rho$.}\label{Fig1}
\end{figure}

The total energy is calculated numerically, and its value is $7.63$ TeV in
qualitative agreement with the energy estimate $~8~ TeV$ obtained in past for the sphaleron
in the Weinberg-Salam model ($\lambda \neq 0$) \cite{klinkmanton84}.
Notice, that expressions for the Higgs vector field $\hat m$ are
different in (\ref{sphII},~\ref{sphIII}),
and the gauge field components $K_3=A_\varphi$ differ in all three solutions.
However, by explicit calculating one can check that expressions for the gauge invariant
Abelian potential ${\cal A}_{m}$, (\ref{calA}), are the same
for all three gauge equivalent representations of the sphaleron
as it should be.
A respective gauge invariant magnetic field (\ref{emtthooft}) has two
non-vanishing magnetic field components
\bea
{\cal F}_{r\theta}&=&0, \nn \\
{\cal F}_{\varphi r}&=& K' \sin^2\theta, \nn \\
{\cal F}_{\theta\varphi}&=&-(1+K) \sin 2 \theta.
\eea

The solution possesses magnetic flux through the plane
$(X,Y)$ which is a multiple of $4 \pi$
\bea
\Phi|_{\theta=\frac{\pi}{2}}&=& \int_0^\infty dr \int_0^{2 \pi} d\varphi {\cal F}_{\varphi r}=-4 \pi.
\eea
The radial vector magnetic field corresponds to the
magnetic field component ${\cal F}_{\theta\varphi}$
shown in Fig. 2.

\begin{figure}[htp]
\centering
\includegraphics[width=60mm,height=55mm]{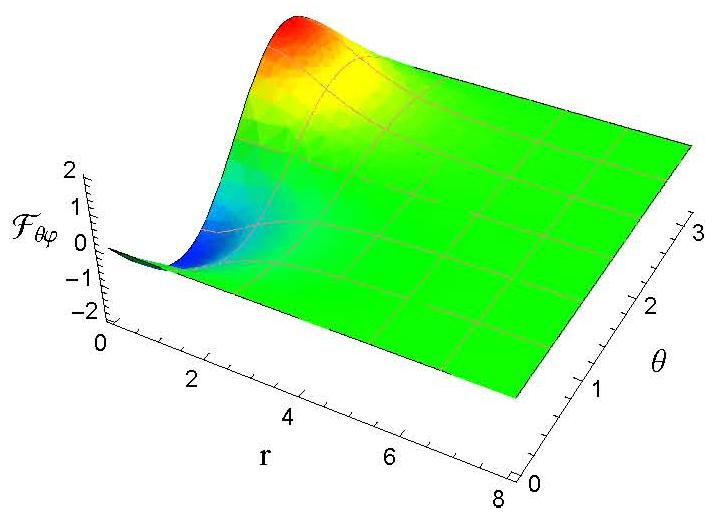}
\caption[plot2]{Magnetic field ${\cal F}_{\theta,\varphi}$.}\label{Fig2}
\end{figure}

The total magnetic flux created by ${\cal F}_{\theta\varphi}$
through any two-dimensional sphere
centered at the origin vanishes identically.
However magnetic fluxes through the upper and lower hemispheres $H^{\pm}$
do not vanish. One can consider magnetic flux through the closed
surface ${\cal S}^\pm$ composed from the upper (lower) hemisphere $H^\pm$ of radius $"a"$
and a disc $D^2:\{r\leq a\}$ in the plane $(X,Y)$. This allows to define magnetic charges
${\cal Q}^\pm_m(a)$ of the monopole and antimonopole
which depend on radius $"a"$
\bea
\Phi_\pm (a)&=& \int_H^\pm d\theta d\varphi {\cal F}_{\theta\varphi}
         \pm \int_{D^2} dr d\varphi {\cal F}_{\varphi r} \nn \\
              &=&4 \pi {\cal Q}^\pm_m(a).
\eea
The magnetic flux through the disc $D^2$ vanishes in the limit
$a\rightarrow 0$. So, one can compactify the hemispheres to spheres
by identifying all points at the boundary of each hemisphere.
Magnetic flux $\Phi_\pm (a)$ through the upper (lower) hemisphere
reaches a maximal value $-4\pi$ ($+4 \pi$) in the limit $a\rightarrow 0$.
With this we obtain the maximal values of the magnetic charges ${\cal Q}_m^{max}=\mp 1$
for the antimonopole and monopole placed in one point $r=0$.
Magnetic fluxes through the upper and lower hemispheres
at space infinity vanish, i.e., each of magnetic charges of the monopole
and antimonopole is totally screened at space infinity.

It is known that total energy density of the sphaleron
is spherically symmetric. However,
the energy corresponding to the gauge invariant Abelian magnetic field
${\cal F}_{mn}$ has axial symmetry. Notice, the energy density
of the Abelian magnetic field is not spherically symmetric,
and near the origin $r\simeq 0$ it is proportional to
\bea
{\cal E}({\cal F}) \simeq \dfrac{1}{r^4} \cos^2 \theta,
\eea
which shows the presence of two relative maximums located
at the points $(r, \theta = 0)$ and $(r, \theta = \pi)$,
i.e., the maximums merge in the limit $r \rightarrow 0$.
The existence of a non-trivial solution for monopole and antimonopole
in the limit when they collapse at one point is an essential feature of
non-linear structure of the non-Abelian theory \cite{plbI}.
Notice, in the case of Abelian theory like Maxwell electrodynamics
the Dirac monopole and antimonopole have mutual attraction and
collapse to a trivial solution.

The sphaleron solution in the Weinberg-Salam model
has a non-vanishing dipole magnetic moment \cite{klinkmanton84}.
To explain this it was suggested in \cite{hind94,hind94b,volkov} to interpret
the sphaleron
as a monopole-antimonopole pair which resembles the
Nambu's ¡°dumb-bell¡± \cite{nambu77}.
Our consideration demonstrates that with a proper
definition of the electromagnetic field
the sphaleron solution represents a monopole and antimonopole placed in one point
at the origin. One should notice, that one can still interpret
the sphaleron as Nambu's "dumb-bell" in the limit of an infinitesimally
small string. In this limit the monopole
and antimonopole become close to each other, and
the system represents a monopole and
antimonopole merged at a single point.
One can verify, that asymptotic behavior of the magnetic field
with our definition is the same as in \cite{klinkmanton84,hind94},
so that sphaleron has the same dipole magnetic moment.
A principal difference from the description in \cite{hind94, hind94b,volkov}
appears in the structure of the sphaleron at small distance, namely,
the magnetic field in our treatment has a singularity at the origin of the type $\dfrac{1}{r^2}$,
whereas the definition for the electromagnetic tensor with (\ref{emthn}) accepted in
\cite{hind94}
leads to regular behavior. One should stress that singularity
of the magnetic field at the origin
does not imply any inconsistences in the theory due to the following reasons.
First of all, the sphaleron solution is regular everywhere and has a finite total energy.
This takes place due to mutual cancelation of the contributions
of the Abelian gauge field and off-diagonal components of $SU(2)$ gauge potential in (\ref{emthn}).
In a fact, the magnetic field of the known 't Hooft-Polyakov monopole
in YMH theory with a real triplet Higgs field has the same singularity
at the origin which does not cause any inconsistencies.
Secondly, the presence of such a singularity in the electromagnetic part
of the classical sphaleron solution
has rather a deep physical origin, and it can be treated as a reflection of
the quantum structure of the Weinberg-Salam model.
It is known that quantum electrodynamics
itself is not a consistent quantum theory due to presence of a positive beta function
which implies existence of the Landau pole in the theory.
Only within unification of the electro-weak interaction
one has a consistent quantum theory
with asymptotically free behavior provided by a negative beta function.
This effect holds due to dominant contribution of $SU(2)$ gauge fields
to the running fine coupling constant in asymptotic regime at high energy scale.
That means, at short distance in quantum description of the sphaleron
one should have cancellation of contributions of the electromagnetic field
and off-diagonal $SU(2)$ gauge bosons. So that, the singularity
of the magnetic field of the classical sphaleron solution
will disappear in quantum description.

\section{On solutions with topological azimuthal magnetic flux}



To find a magnetic solution with a non-vanishing azimuthal magnetic flux
having topological origin one should
apply a general axially-symmetric ansatz and solve fifteen equations
of motion of the Weinberg-Salam model which is a hard problem.
We suppose that such a solution exists and study its properties
using variational method.
For simplicity we restrict our consideration by the case of pure Yang-Mills theory.
In this case one can find a restricted ansatz
which admits local solutions to all equations of motion near the origin, $r\simeq 0$,
and in asymptotic region, $r\simeq \infty$.
Using these solutions we apply variational procedure
to find field configuration minimizing the energy functional.

First we provide a qualitative argument to existence
of a magnetic solution with a magnetic flux around $Z-$ axis.
Let us consider a simple $CP^1$ type model defined by
the energy density obtained from (\ref{energy}) by
keeping only terms with the Higgs field components $\hat m$ and $\rho$
\bea
E_1&=&\dfrac{1}{4}H_{mn}^2+\dfrac{1}{2}\vec C_m^2 \rho^2, \label{CP1en}
\eea
where $\vec C_m$ is defined by (\ref{abeldec}) with
$\hn \equiv \hat m$.
For simplicity we replace the Higgs scalar field $\rho$ with
its classical vacuum averaged value, $\rho=1$.
In this approximation the energy density (\ref{CP1en})
defines a modified Skyrme model which admits exact finite
energy solutions \cite{ferreira13}.
Simple scaling arguments based on the Derrick theorem \cite{derrick}
imply that such a model admits stable static solitons.
We parameterize the $CP^1$ field $\hat m$ with only one function
$Q$ whereas setting the function $f$ to a constant value, $f\equiv 1$.
Changing variable,
$Q=\cot(\dfrac{S}{2})$, one can simplify the Euler-Lagrange equation of motion
for $S(r,\theta)$
\bea
&&(r^2 \cos^2\theta+4\sin^2\theta \sin^2\dfrac{S}{2})S_{rr}+
(\dfrac{1}{r^2} \sin^2\theta\sin^2S + \nn \\
&& \cos^2\theta)S_{\theta\theta}+\dfrac{1}{r^2} \sin^2\theta\sin S\big(\cos S S^2_\theta
-4 \sin^2\dfrac{S}{2}\nn+ \\
&&3 \cot\theta\sin S S_\theta \big )
+ \dfrac{1}{4 \sin\theta} (\cos\theta+3\cos(3\theta)) S_\theta- \nn \\
&&2 \cos^2\theta(\sin S-r S_r)+\sin^2\theta \sin S S_r =0.
\eea
 Finite energy condition allows the following boundary conditions
\bea
S(0,\theta)&=&0, ~~~~~S(\infty, \theta)=2 \pi
\eea
which provide a total magnetic flux $4 \pi$ for
the azimuthal magnetic field through the half plane
 $\{y=0; x \geq 0\}$.
A regular numeric solution is obtained by using the package COMSOL
for solving partial differential equations.
The results for the function $S$ and energy density ${\cal E}^{CP^1}$
are presented in Figs. 3, 4.
\begin{figure}[htp]
\centering
\includegraphics[width=65mm,height=45mm]{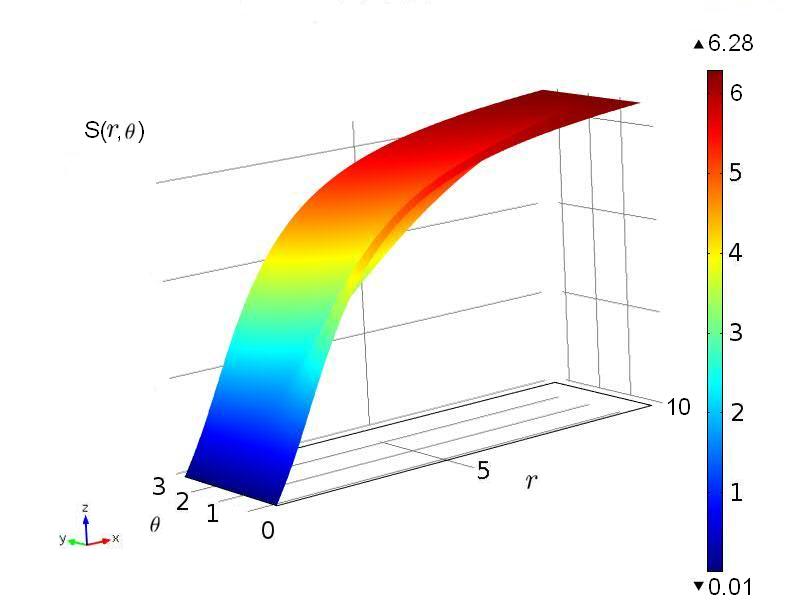}
\caption[plot3]{Solution for $S(r,\theta)$. }\label{Fig3}
\end{figure}
The total energy is $~18$ TeV and the contribution of the first term in (\ref{CP1en})
is 51 \% of the total energy in agreement with the estimate
50 \% based on Derrick theorem.
Notice that homogeneous configurations of the
Higgs field $\rho$ and $f$ do not represent
real solutions to the original equations of motion of the Weinberg-Salam model.
We suppose  that a real magnetic solution with the helical structure
might include non-trivial field configurations for all field variables
within the general axially-symmetric ansatz.

Let us consider now the Yang-Mills-Higgs theory.
We choose a parametrization (\ref{higgspar}) for the Higgs complex doublet
with an axially-symmetric scalar field $\rho (r,\theta)$ ($\omega(x)=0$)
and a complex scalar doublet $\tilde \zeta$ determined by the ansatz (\ref{ansfQ})
with $\hat m\equiv \hn$ and winding numbers $m=1, n=2$.
We impose a constraint $f(r,\theta)\equiv 1$ and the following boundary conditions for
the function $Q(r,\theta)$
\bea
Q(0,\theta)=+\infty, ~~~~Q(\infty,\theta)=0.
\eea
At space infinity the Higgs triplet
vector field $\hat m$ takes asymptotic form
\bea
\hat m&=&\Big (\cos \varphi \sin (2 \theta),~\sin \varphi \sin(2 \theta),~ \cos (2 \theta) \Big).
\eea
We decompose the gauge potential in the natural basis frame
\bea
\vec A_m(r,\theta) &=& A_m(r,\theta) \hat m-\hat m\times \der_m \hat m ~W(r,\theta). \label{helicalans}
\eea
Notice, in the case when the off-diagonal component $W(r,\theta)$ equals one,
one has an ansatz with five functions $A_m, Q, \rho$ which reduces all equations
of motion to five independent second order equations.
This ansatz might be too restricted since by $SU(2)$ gauge transformation
one can rotate the vector field $\hat m$ to a constant unit vector $(0,0,1)$
everywhere. With this the Lagrangian of YMH theory becomes formally
equivalent to the Lagrangian of the Ginzburg-Landau
model with a complex scalar field in the unitary gauge.
The ansatz (\ref{helicalans}) with the constraint $W=1$
can not admit finite energy solutions with a non-zero azimuthal magnetic flux,
so we keep a non-constant off-diagonal gauge field $W(r,\theta)$.
\begin{figure}[htp]
\centering
\includegraphics[width=65mm,height=45mm]{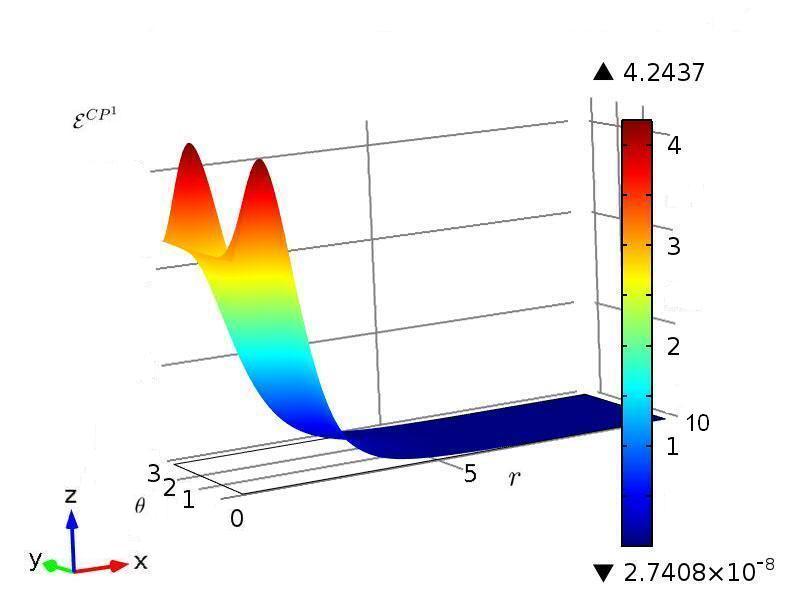}
\caption[plot4]{Energy density plot.}\label{Fig4}
\end{figure}

Direct substitution of the ansatz (\ref{helicalans})
into all equations of motion leads to number of independent equations
more than number of unknown variables.
We have solved all equations of motion near the origin and in asymptotic region
in the gauge $A_r=0$
and obtain local solutions which are consistent with finite energy condition.
The solution at the origin $r\simeq 0$ can be found by perturbation
theory by series expansion as follows
\bea
&&G(r,\theta)=C_1 r- C_1 r^2 + \dfrac{C_1}{240 \rho_0} r^3 \Big (120 C_2+ \nn \\
&&~~~~~\rho_0\big(-37 \rho_0^2+4C_1^2(w_0(w_0-2)+60)\big)-\nn \\
&&~~~~~15\big(24C_2-\rho_0(5 \rho_0^2-12C_1^2w_0(w_0-2))\big )\cos^2\theta\Big), \nn \\
&&W(r,\theta)=w_0+\dfrac{w_0-1}{48}r^2 \Big(3\big(5\rho_0^2-12C_1^2w_0(w_0-2)\big) \nn \\
&&~~~~~+\big(-7 \rho_0^2 +4 C_2^2w_0 ( w_0-2) \big ) \cos(2\theta \Big ), \nn \\
&&A_\theta(r,\theta)=\dfrac{C_1}{3} r^3 \big(-\rho_0^2+4 C_1^2 w_0(w_0-2) \big)\sin\theta,\nn \\
&&A_\varphi(r,\theta)=-\dfrac{1}{12}r^2 \big (-\rho_0^2+28 C_1^2 w_0(w_0-2) \big)\sin^2\theta ,\nn \\
&&\rho(r,\theta)=\rho_0+\dfrac{1}{24} r^2 \Big( 6 C_2 +4 C_1^2 \rho_0 \big (3+2 w_0(w_0-2)\big) \nn \\
&&~~~~~+18 C_2 \cos (2 \theta) \Big),
  \eea
where $w_0, \rho_0$ are free variational parameters
for initial values of $W, \rho$ at the origin.
We have performed a change of variables
 $Q(r,\theta)\rightarrow G(r,\theta)=(1+Q(r,\theta))^{-1}$
which is suitable for numeric purpose.
The solution contains only those independent integration constants which
provide finite energy conditions and the symmetry under the reflection
$(z\rightarrow -z)$.
In a similar manner one can find a solution in asymptotic space region $r\simeq \infty$.
The solution includes an essential singularity which can
be extracted in the exponential factor
\bea
G(r,\theta)&=& 1+\tC_1 \Big (\dfrac{e^{-r}}{r^2}(1+r)+\text{Chi}(r)-\mbox{Shi}(r)\Big ), \nn \\
W(r,\theta)&=&1- \tC_2 \dfrac{e^{-r}}{r}(1+r), \nn \\
A_\theta(r,\theta)&=&-2\tC_1\Big(\dfrac{e^{-r}}{r^2}(2+2r+r^2) \nn \\
 &&+\mbox{Chi(r)}-\mbox{Shi(r)}\Big)\sin\theta, \nn\\
A_\varphi(r,\theta)&=&\Big(2+\tC_3 \dfrac{e^{-r}}{r}(1+r)\Big) \sin^2 \theta , \nn\\
\rho(r,\theta)&=&1+\dfrac{\tC_4}{r}+\dfrac{\tC_5}{r^3} (1+3 \cos (2 \theta)),
\eea
where $\mbox{Chi(r)},\mbox{Shi(r)}$ are the special cosine and sine
hyperbolic integral functions,
and for simplicity the Higgs coupling constant $\lambda$ is set to zero.
If we take into account a non-zero value for the coupling constant $\lambda$
then the Higgs scalar $\rho$ will approach its asymptotic value in
exponential form as well.
We will apply variational method to estimate the energy of the solution.
To simplify variational procedure one can observe, that
the minimum of the energy requires a simple condition
for the azimuthal component of the gauge potential $A_\varphi(r,\theta)$, namely
\bea
A_\varphi(r,\theta)=-\tilde C_\varphi (r,\theta)=\dfrac{2 \sin^2 \theta}{1+Q^2}. \label{Aficond}
\eea
The condition (\ref{Aficond}) provides
mutual cancelation of contributions to the energy
from the fields $H_{r\phi}, H_{\theta \phi}$
and $F_{r\phi}, F_{\theta \phi}$ respectively as it can be seen from
the Abelian structure of the field strength in (\ref{cdual}).
With this we construct trial variational functions
for the fields $(G(r,\theta), A_{\theta,\varphi} (r,\theta),
W(r,\theta),\rho(r,\theta))$ in a consistent manner with the
known local solutions taking into account first derivative terms as well.
To make a qualitative estimate we consider radial dependent functions $G(r),\rho(r)$
and factorize the angle dependence of the gauge potentials $A_{\theta,\varphi} (r,\theta)$
as follows
\bea
A_\theta (r,\theta)&=&a_\theta (r) \sin\theta, \nn \\
A_\varphi (r,\theta)&=&a_\varphi(r) \sin^2\theta,
\eea
where $a_\varphi (r)$ is determined by equation (\ref{Aficond}).
Minimization procedure of the energy functional
produces an upper bound for the total energy,
 $E_{bound}\simeq 4.65$ TeV.
Variational profile functions $W, G, a_\theta, a_\varphi,\rho$ minimizing
the energy are presented in Fig. 5.
\begin{figure}[htp]
\centering
\includegraphics[width=60mm,height=55mm]{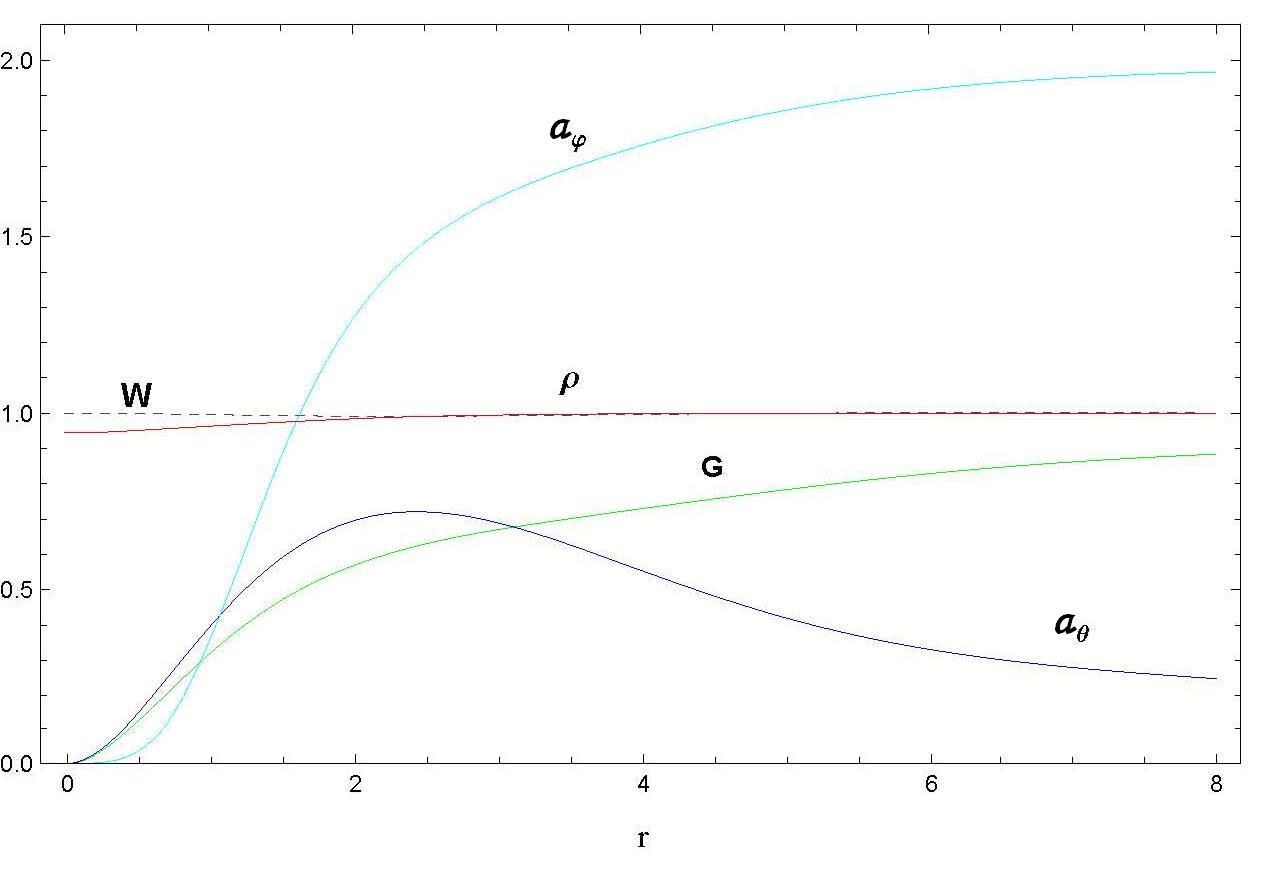}
\caption[plot5]{Variational profiles for the functions $G,W, a_\theta, a_\varphi,\rho$. }
\label{Fig5}
\end{figure}
A gauge invariant Abelian magnetic field, ${\cal F}_{mn}$,
has only one non-vanishing component, ${\cal F}_{r \theta}$,
which provides a non-zero magnetic flux around the $Z$-axis
\bea
\Phi_\varphi&=& \int dr d\theta {\cal F}_{r\theta}=\int dr d\theta H_{r\theta}=2 \pi.
\eea
One can extract the energy density
${\cal E}({\cal F})=\dfrac{1}{4}{\cal F}_{mn}^2$ corresponding to the
Abelian magnetic field ${\cal F}_{mn}$,
(\ref{emtthooft}). Note that the respective total energy
is finite, $E_{magn}=1.58$ TeV, contrary to the case of the sphaleron solution.

One should stress, due to additive structure of the $SU(2)$ gauge invariant
Abelian field strength ${\cal F}_{\mu\nu}$, (\ref{emtthooft}),
one has partial mutual cancelation of contributions
of the fields $F_{\mu\nu}$ and $H_{\mu\nu}$.
Without such cancelation, i.e., if we neglected
the field $F_{\mu\nu}$, we would have energy
contribution of the Higgs field provided by the
term $\frac{1}{4}H_{mn}^2$ about
4.9 TeV what is three times larger than the actual energy
$1.58$ TeV of the Abelian gauge field ${\cal F}_{\mu\nu}$.
So that interaction between the Higgs
field and $SU(2)$ gauge bosons leads to significant decrease of the
Abelian magnetic field energy and provides
magnetic charge screening effect in the case of monopole like solutions.

\section{Conclusion}

    We propose a most general ansatz for static axially-symmetric
magnetic field configurations in the Weinberg-Salam model based on gauge invariant
decomposition of $SU(2)$ gauge potential.
Introducing a unique $SU(2)$ gauge invariant definition for the
electromagnetic vector gauge potential we have proved that
for any possible finite energy monopole like
solutions the magnetic charge of monopole (antimonopole)
must be totally screened at space infinity.
Our analysis implies that finite energy condition and equations of motion
forbid existence of solutions representing a system of
monopoles and antimonopoles localized in different points.
We have demonstrated that known sphaleron solution represents a pair
of monopole and antimonopole placed in one point.
In general, an axially-symmetric magnetic field configuration
can possess a helical magnetic structure. We conjecture that such a solution
may exist in the standard electroweak  model and describe a possible simple
solution with azimuthal magnetic flux
using variational method in the case of a pure Yang-Mills Higgs theory.
The solution possesses a total magnetic flux $2 \pi$
around the $Z$-axis.
We estimate the energy of the solution, $E=4.3$ TeV.
In the case of the Weinberg-Salam model the energy of such magnetic
solution is expected to be of the same order as the energy of sphaleron.
So the sphaleron and the magnetic solution with aziumthal magnetic flux
could be good candidates in search of magnetic like
bound states in the electroweak theory.
\acknowledgments
Authors thank J. Evslin for numerous useful comments
and E. Tsoy for stimulating discussions.
The work is supported by NSFC (Grants 11035006 and 11175215),
the Chinese Academy of Sciences visiting professorship for senior international
scientists (Grant No. 2011T1J3), and by UzFFR (Grant F2-FA-F116).

\end{document}